 \documentclass[final,5p,times,twocolumn]{elsarticle}

 \usepackage{graphicx}
 \usepackage{epsfig}

\usepackage{amssymb}
\usepackage{amsthm}
\usepackage{amsmath, array, bm}
\usepackage{lineno}
\usepackage{hyperref}
\usepackage{listings}
\usepackage{xcolor}
\usepackage{threeparttable}
\usepackage{booktabs}
\usepackage{multicol, blindtext}
\usepackage[ruled,vlined]{algorithm2e}
\usepackage{dirtytalk}
\definecolor{commentgreen}{RGB}{2,112,10}
\definecolor{lightgray}{gray}{0.9}
\newcounter{bla}

\journal{Computer Physics Communications}

\begin{document}
\begin{frontmatter}



\title{JRAF: A Julia Package for Computation of the Relativistic Molecular Auxiliary Functions}


\author{Ali Ba\u{g}c{\i}\corref{author}}

\cortext[author] {Corresponding author.\\\textit{E-mail address:} abagci@pau.edu.tr}
\address{Department of Physics, Faculty of Arts and Sciences, Pamukkale University, \c{C}amlaraltı, K{\i}n{\i}kl{\i} Campus, Denizli, Turkey}

\begin{abstract}
Evaluation of relativistic molecular integrals over exponential$-$type spinor orbitals require using the relativistic auxiliary functions in prolate spheroidal coordinates. They have derived recently by the author [Physical Review E \textbf{91}, 023303 (2015)]. They are used in solution of the molecular Dirac equation for electrons moving around Coulomb potential. A series of papers on a method for fully analytical evaluation of relativistic auxiliary functions in following, published \cite{2_Bagci_2018, 3_Bagci_2018, 4_Bagci_2020}. From the computational physics point of view, these works also demonstrate how to deal with the integrals involve product of power functions with non$-$integer exponents and incomplete gamma functions. The computer program package to calculate these auxiliary functions in high accuracy is presented. It is designed in $Julia$ programming language. It is capable of yielding highly accurate results for molecular integrals over a wide range of orbital parameters and quantum numbers. Additionally, the program package provides numerous tools such as efficient calculation for the angular momentum coefficients arising in the product of two normalized associated Legendre functions centered on different atomic positions, the rotation angular functions used for both complex and real spherical harmonics. Sample calculations are performed for two$-$center one$-$electron integrals over non$-$integer Slater$-$type orbitals. The results prove that the package is robust. 
\end{abstract}


\end{frontmatter}


\pagebreak
{\bf PROGRAM SUMMARY}

\begin{small}
\noindent
{\em Program Title:} JRAF                                         		  				\\
{\em CPC Library link to program files:} (to be added by Technical Editor)  				\\
{\em Developer's repository link:} \url{https://github.com/abagciphys/JRAF.jl} 			\\
{\em Code Ocean capsule:} (to be added by Technical Editor)				  				\\
{\em Licensing provisions:} MIT				  				\\
{\em Programming language:} $Julia$ Programming Language                      				\\
{\em Supplementary material:} An experimental version of the computer program package written in $Mathematica$ Programming Language \cite{5_Mathematica}.						 						  	\\
{\em External routines/libraries:} $Nemo$ computer algebra package for the $Julia$ programming language \cite{6_Fieker_2017}, $Cuba$ multidimensional numerical integration with different algorithms in $Julia$ \cite{7_Hahn_2007}.									             	\\
{\em Nature of problem:} The relativistic molecular auxiliary functions integrals are the results of the expression for two$-$center two$-$electron Coulomb energy associated with a charge density. The Coulomb energy is transformed into a kinetic energy integrals using the Poisson's equation and single$-$center potential, considering the Laplace expansion for Coulomb interaction are expressed in terms of normalized non$-$integer Slater$-$type orbitals \cite{1_Bagci_2015}. 
Using the resulting expression into the two$-$center two$-$electron integrals the relativistic auxiliary functions integrals are derived in the prolate ellipsoidal coordinates. These auxiliary functions in the following, generalized to the whole set of physical potential operators through an argument such that the Coulomb potential particular case of it.\\
The relativistic auxiliary functions integrals have no closed$-$form solutions except that their parameters are in the set of integer numbers. This makes analytical evaluation of the relativistic molecular auxiliary functions challenging. They are used in solution of matrix form representation for the molecular Dirac$-$Fock self$-$consistent field (SCF) equation. \\
{\em Solution method :} A criterion considering the symmetry properties of two$-$center two$-$electron molecular integrals is proposed at first \cite{2_Bagci_2018}. This lead to avoid the computation of the incomplete and the complementary incomplete gamma functions and take advantage of the sum between them $\left(P+Q=1\right)$. The resulting form of the relativistic molecular auxiliary functions integrals are expressed in terms of functions involving convergent series representation for incomplete beta functions. Recurrence relationships are then, derived for each of these sub$-$functions \cite{3_Bagci_2018}. The algorithm of the code written for computation of auxiliary functions is based on the vectorization procedure defined in \cite{4_Bagci_2020}.

\end{small}

\section{Introduction}\label{Intro}
The solution for the matrix form representation of molecular Dirac equation requires using exponential$-$type spinors basis if the nuclei considered as a point$-$like \cite{1_Shabad_2005, 2_Gitman_2013, 3_Schwerdtfeger_2015}. To avoid the \textit{variational collapse} \cite{4_Schwarz_1982, 5_Schwarz_1982}, these kinds of spinors basis have to be derived from the $L-$spinors \cite{6_Quiney_1989, 7_Grant_2007},
\begin{linenomath*}
\begin{multline}\label{eq:1}
f^{\beta}_{n_r\kappa}(\zeta, r)=
\left[\frac{n_{r}! (2\gamma+n_{r})}{2N_{n_{r}\kappa}(N_{n_{r}\kappa}-\kappa)\Gamma(2\gamma+n_{r})} \right]
{(2\zeta r)^{\gamma}}{e^{-\zeta r}}
\\
\times \left\lbrace -(1-\delta_{{n_r}0})L^{2\gamma}_{n_{r}-1} (2\zeta r)
+\beta \left(\frac{N_{n_{r}\kappa}-\kappa}{n_{r}+2\gamma} \right) {L^{2\gamma}_{n_r}(2\zeta r)} \right\rbrace .
\end{multline}
\end{linenomath*}
They are the solution of the following Dirac$-$Coulomb differential equation defined for hydrogen$-$like atoms \cite{7_Grant_2007, 8_Bostock_2011},
\begin{linenomath*}
\begin{multline}\label{eq:2}
\frac{\partial}{\partial r}f_{n_{r}\kappa}^{\beta}\left(\zeta, r\right)
=-\beta\frac{\kappa}{r}f_{n_{r}\kappa}^{\beta}\left(\zeta, r\right)\\
+\left(\frac{\beta N_{n_{r}\kappa}-\gamma-n_{r}}{r}+ \zeta \right)f_{n_{r}\kappa}^{-\beta}\left(\zeta, r\right),
\end{multline}
\end{linenomath*}
here, $L_{q}^{p}\left(x\right)$ are generalized Laguerre polynimoals. $n_{r}$ is the redial quantum number, with $n_{r}=n-\vert \kappa \vert$, and $n$ the principal quantum number. $\kappa=\left\lbrace \pm1, \pm2, ...\right\rbrace$, are the eigenvalues of an operator commute with Dirac Hamiltonian, $-\beta\left(\bm{\sigma} . \bold{L}+\bold{I} \right)$, where $\bold{L}$ is the orbital angular momentum operator, $\bm{\sigma}=\left\lbrace \sigma_{1}, \sigma_{2}, \sigma_{3} \right\rbrace$, $\sigma_{i}$ are the Pauli spin matrices, $\bold{I}$ denotes $2\times2$ identity matrix, $\beta=\pm 1$ represent $Large-$ and $Small-$components of the wave$-$function, respectively. This nomenclature is preferred for positive energy solution. The $Small-$component go to zero in the non$-$relativistic limit and the $Large-$component become a solution of the corresponding non$-$relativistic equation, i.e. the Schr{\"o}dinger equation. The exponent of the power functions $\gamma$;
\begin{linenomath*}
\begin{align}\label{eq:3}
\gamma=\sqrt{\kappa^2-\frac{Z^2}{c^2}}.
\end{align}
\end{linenomath*}
$Z$ is the atomic number, $c$ is the speed of light. And $N_{n_{r}\kappa}$,
\begin{linenomath*}
\begin{align}\label{eq:4}
N_{n_{r}\kappa}
=\sqrt{n_{r}^{2}+2n_{r}\gamma+\kappa^2}.
\end{align}
\end{linenomath*}
Thus, the four$-$component form of the Dirac equation for a central Coulomb potential becomes free from spin$-$angular part and reduce to solve a system of differential equation given in the Eq. (\ref{eq:2}) which its solution [the Eq. (\ref{eq:1})] since directly coupled, the radial two$-$component spinor naturally ensure that the \textit{kinetic$-$balance} condition is fulfilled. \\
The problem in basis spinors derived from the Eqs. (\ref{eq:1}, \ref{eq:2}) such as $S-$spinors \cite{7_Grant_2007}, Slater$-$type spinors orbitals \cite{9_Bagci_2016} are that they do not pose an addition theorem \cite{10_Biedenharn_1984}. The power function $r^{\gamma}$ in the Eq. (\ref{eq:1}) is non$-$analytic. This is due to that the exponent $\gamma$ is in the set of real numbers. Meaningful power series about $r=0$ are not available. Consequently, obtaining compact form relations for relativistic molecular integrals becomes laborious \cite{11_Weniger_2008, 12_Weniger_2012} (Please see also references therein). Yet, a recently suggested method by author in \cite{13_Bagci_2014} to treat the molecular integrals numerically is able to generate successful results in all range or parameters. In this method, the molecular integrals are expressed in terms of new relativistic molecular auxiliary functions, derived in \cite{14_Bagci_2015}. They are calculated using the numerical Global adaptive method with Gauss$-$Kronrod numerical integration extension \cite{15_Davis_1975, 16_Mathematica_NInteg}.

The relativistic molecular auxiliary functions integrals derived following the procedure given in \cite{17_Weatherford_2005, 18_Weatherford_2006} for solution of the Poisson's equation using spectral forms \cite{19_Weatherford_1985} (and reference therein) where the Coulomb energy were written as a kinetic energy$-$like integral using the Green's theorem. The potential in the resulting expression satisfies Poisson's equation. It was solved as a partial differential equation in spherical coordinates. The potential expanded in the new set of the functions, called spectral forms, which involve the incomplete gamma functions. Series representation for the incomplete gamma functions are computationally unstable \cite{20_Gill_2012, 21_Backeljauw_2014, 22_Jameson_2016, 23_Nemes_2019}. The convergence can be quite slow depending on the values of parameters. This leads to impose some restrictions on the set of functions to be used in the expansion.

The above procedure by the author improved to a general set of functions where the parameters are free from any restriction \cite{14_Bagci_2015}. A symmetry feature for two$-$center two$-$electron molecular integrals that pointed out and identified by a criterion in \cite{24_Bagci_2018, 25_Bagci_2020}, eliminates the necessity for immediate expansion for incomplete gamma functions or using the relations for the normalized complementary incomplete and the normalized incomplete gamma functions as $P=Q-1$, $Q=P-1$. Their conditional convergence \cite{20_Gill_2012, 22_Jameson_2016} cause some restrictions for the set of functions to be used.

\vspace{2mm}
\textbf{Criterion.} Let $P\left[n_{4}-n_{1}, z \right]$ and $Q \left[n'_{4}-n'_{1}, z \right]$; then $n_{4}-n_{1}=a \pm c$, $n'_{4}-n'_{1}=a \pm d$, where $a \in \mathbb{R}$, $\left\lbrace c, d \right\rbrace \in \mathbb{Z}$ holds.
\\
\vspace{2mm}\\
Finally, the following relationship for relativistic molecular auxiliary functions integrals (RAF) obtained in prolate spheroidal coordinates,
\begin{linenomath*}
\begin{multline} \label{eq:5}
\left\lbrace \begin{array}{cc}
\mathcal{P}^{n_1,q}_{n_{2}n_{3}n_{4}}\left(p_{123} \right)
\\
\mathcal{Q}^{n_1,q}_{n_{2}n_{3}n_{4}}\left(p_{123} \right)
\end{array} \right\rbrace\\
=\frac{p_{1}^{\sl n_{1}}}{\left({\sl n_{4}}-{\sl n_{1}} \right)_{\sl n_{1}}}
 \int_{1}^{\infty}\int_{-1}^{1}{\left(\xi\nu \right)^{q}\left(\xi+\nu \right)^{\sl n_{2}}\left(\xi-\nu \right)^{\sl n_{3}}}\\ \times
\left\lbrace \begin{array}{cc}
P\left[{\sl n_{4}-n_{1}},p_{1}f_{ij}^{k}\left(\xi, \nu\right) \right]
\\
Q\left[{\sl n_{4}-n_{1}},p_{1}f_{ij}^{k}\left(\xi, \nu\right) \right]
\end{array} \right\rbrace
e^{p_{2}\xi-p_{3}\nu}d\xi d\nu,
\end{multline}
\end{linenomath*}
where,
\begin{linenomath*}
\begin{align}\label{eq:6}
f_{ij}^{k}\left(\xi, \mu \right)
=\left(\xi \mu\right)^{k}\left(\xi+\nu \right)^{i}\left(\xi-\nu\right)^{j},
\end{align}
\end{linenomath*}
stands to represent the elements required to generate a potential. For Coulomb potential it has a form that, $i=1$, $k=j=0$; $f_{10}^{0}\left(\xi, \nu\right)=\left(\xi+\nu \right)$. Note that, for any values of $f_{ij}^{k}$ the identified criterion is still valid. This means that an analytical method of solution to be found for the Eq. (\ref{eq:5}) considering the Coulomb potential is available for any other type of potential. The relativistic molecular auxiliary functions integrals are analogous to radial two$-$component spinor. They are the results of two$-$center two$-$electron interaction. The spherical symmetry only contribute with a term that $\left(\xi \nu\right)^{q}$ which easily be eliminated since $q \in Z$. They have two$-$component form. For arbitrarily potential thus, validity of the criterion given about leads to represent the two$-$electron interaction in terms of one$-$electron interaction.

A computer program based on the analytical methods of computation, described in the previous series of papers \cite{24_Bagci_2018, 26_Bagci_2018, 27_Bagci_2020} by the author for relativistic molecular auxiliary functions, is presented here. It is unique inasmuch as it is the only computer program code that enables highly accurate calculation of the molecular integrals involving power functions with non$-$integer exponents. It has been successful among the enormous number of attempts. The history of non$-$integer principal quantum number usage comprehensively discussed in \cite{27_Bagci_2020}.

In Section \ref{MSolution} we briefly recall the analytical method of solution for the seek of completeness. Details of implementation is discussed in the Section \ref{DJRAF}. This section is also devoted to description of the $JRAF$ package including its features, details about usage and a comprehensive test run. In the Section \ref{ResDis} we conclude about efficiency of the computer program package by presenting benchmark results of molecular auxiliary functions and some two$-$center molecular integrals involving them. A code written by the author in $Mathematica$ programming language \cite{28_Mathematica} to perform the calculations using the numerical Global$-$adaptive strategy is considered for comparison.
\section{A Convergent Series Representation for RAF}\label{MSolution}
According to the criterion given in the previous section and the following property for the normalized incomplete gamma functions,
\begin{linenomath*}
\begin{align}\label{eq:7}
P\left[a,z \right]=\frac{\gamma \left(a,z \right)}{\Gamma\left(z\right)},
\hspace{4mm}
Q\left[a,z \right]=\frac{\Gamma \left(a,z \right)}{\Gamma\left(z\right)},
\hspace{4mm}
P+Q=1,
\end{align}
\end{linenomath*}
the problem of evaluation for relativistic molecular auxiliary functions reduce to the following form;
\begin{linenomath*}
\begin{align}\label{eq:8}
\mathcal{P}^{n_1,q}_{n_{2}n_{3}n_{4}}\left(p_{123} \right)
+
\mathcal{Q}^{n_1,q}_{n_{2}n_{3}n_{4}}\left(p_{123} \right)
=
\mathcal{G}^{n_1,q}_{n_{2}n_{3}n_{4}}\left(p_{123} \right),
\end{align}
\begin{multline}\label{eq:9}
\mathcal{G}^{n_1,q}_{n_{2}n_{3}n_{4}}\left(p_{123} \right)
=
\frac{p_{1}^{\sl n_{1}}}{\left({\sl n_{4}}-{\sl n_{1}} \right)_{\sl n_{1}}}
\\
\times \int_{1}^{\infty}\int_{-1}^{1}{\left(\xi\nu \right)^{q}\left(\xi+\nu \right)^{\sl n_{2}}\left(\xi-\nu \right)^{\sl n_{3}}}
e^{p_{2}\xi-p_{3}\nu}d\xi d\nu.
\end{multline}
\end{linenomath*}
The integrals given in the Eq. (\ref{eq:9}) are also sum of two integrals involving Appell's functions. 
\begin{linenomath*}
\begin{multline}\label{eq:10}
{\mathcal{G}^{n_{1},q}_{n_{2}n_{3}}}(p_{123})
=\frac{p_{1}^{n_{1}}}{\Gamma\left(n_{1}+1 \right)}
\sum_{s=0}^{\infty}
\frac{p_{3}^{s}}{\Gamma\left(s+1 \right)}
\left(\frac{1}{q+s+1}\right)
\\
\times
\left\lbrace
J_{n_{2}n_{3}}^{q+s+1,q+s+2;q}\left(p_{2}\right)
+\left(-1\right)^{s}J_{n_{3}n_{2}}^{q+s+1,q+s+2;q}\left(p_{2}\right)
\right\rbrace,
\end{multline}
\end{linenomath*}
where, 
\begin{linenomath*}
\begin{multline}\label{eq:11}
J_{n_{2}n_{3}}^{s,s';q}\left(p\right)\\
=\int_{1}^{\infty}F_{1}\left(s;-n_{2},-n_{3};s';\frac{1}{\xi},-\frac{1}{\xi} \right)
\xi^{n_{2}+n_{3}+q}e^{-p\xi}d\xi.
\end{multline}
\end{linenomath*}
The definition for Appell's hypergeometric functions used here \cite{29_Appell_1925},
\begin{linenomath*}
\begin{multline}\label{eq:12}
F_{1}\left(\alpha;\beta_{1},\beta_{2};\gamma;x,y \right)=
\frac{\Gamma\left(\gamma\right)}{\Gamma\left(\alpha\right)\Gamma\left(\alpha-\gamma\right)}\\
\times \int_{0}^{1}u^{\alpha-1}\left(1-u\right)^{\gamma-\alpha-1}
\left(1-ux\right)^{-\beta_{1}}
\left(1-uy\right)^{-\beta_{2}}du.
\end{multline}
\end{linenomath*}
The analytical expression to be used for computation of the Eq. (\ref{eq:9}) based on the vectorization procedure, were suggested explicitly in \cite{27_Bagci_2020} as,
\begin{linenomath*}
\begin{multline}\label{eq:13}
G^{n_{1},q}_{n_{2},n_{3}}\left(p_{123} \right)\\
=\frac{p_{1}^{n_{1}}}{\Gamma\left(n_{1}+1\right)}
\frac{1}{2^{2q}}
\sum_{s_{1},s_{2},s_{3}} \left(-1\right)^{s_{1}+s_{2}+s_{3}}F_{s_{1}}\left(q\right)
\frac{1}{2^{s_{2}}}
 F_{s_{3}}\left(s_{2}\right)
 \\
\times \left\lbrace
\frac{1}{2^{-s_{2}}}
\left(
\frac{p_{3}^{s_{2}}}{\Gamma\left(s_{2}+1\right)}
2^{n_{2}+n_{3}+2q+s_{2}+1}
B_{n_{2}+2q-2s_{1}+2s_{2}-2s_{3}+1, n_{3}+2s_{1}+2s_{3}+1}
\right.
\right.
\\
\times E_{-\left(n_{2}+n_{3}+2q+s_{2}+1 \right)}\left(p_{2}\right)
-l^{s_{2},-s_{2}}_{n_{2}+2q-2s_{1}+2s_{2}-2s_{3}, n_{3}+2s_{1}+2s_{2}}\left(p_{302}\right)
\\
-l^{s_{2},-s_{2}}_{n_{3}+2s_{1}+2s_{2}, n_{2}+2q-2s_{1}+2s_{2}-2s_{3}}\left(p_{302}\right)
\left.
\left.
\bigg) \right.
\right\rbrace.
\end{multline}
\end{linenomath*}
$0 \leq s_{1} \leq q$, $0 \leq s_{2} \leq N$, $0 \leq s_{3}, \leq s_{2}$, $N$ is used to indicate the upper limit of summation. $B_{n,n'}$ are the beta functions. Note that, the vectorization procedure runs faster than the corresponding code contaninig loops. The Eq. (\ref{eq:13}) contains four indices one of them $\left( s_{4} \right)$ is in a sub$-$function belongs to $l^{n_{1},q}_{n_{2},n_{3}}\left(p \right)$ auxiliary function.
\begin{linenomath*}
\begin{multline}\label{eq:14}
l^{n_{1},q_{1}}_{n_{2}n_{3}}\left(p_{123}\right)
=\frac{p_{1}^{n_{1}}}{\Gamma\left(n_{1}+1 \right)}e^{-p_{2}}
\\
\times \int_{1}^{\infty}\left(2\xi \right)^{n_{2}+n_{3}-q_{1}+1}B_{n_{2}+1,n_{3}+1}\left(\frac{\xi+1}{2\xi} \right) e^{-p_{3}\xi}d\xi,
\end{multline}
\begin{multline}\label{eq:15}
l^{n_{1},q_{1}}_{n_{2}n_{3}}\left(p_{123}\right)
=\frac{p_{1}^{n_{1}}}{\Gamma\left(n_{1}+1\right)}e^{-p_{2}} \\
\times \sum_{s_{4}}\frac{\left(-n_{2}\right)_{s_{4}}}{\left(n_{3}+s_{4}+1\right)s_{4}!}m^{n_{2}+q_{1}-s_{4}}_{n_{3}+s_{4}+1}\left(p_{3}\right),
\end{multline}
\end{linenomath*}
here, $B_{n,n'}\left(z \right)$ are the incomplete beta functions \cite{30_Abramowitz_1974} and,
\begin{linenomath*}
\begin{align}\label{eq:16}
m^{n_{1}}_{n_{2}}\left(p\right)
=2^{n_{1}}e^{-p}
\int_{0}^{\infty} \left(1+v\right)^{n_{1}} v^{n_{2}} e^{-p v}dv.
\end{align}
\end{linenomath*}
The integral on the left$-$hand side is the confluent hyper$-$geometric function of the second kind \cite{30_Abramowitz_1974},
\begin{linenomath*}
\begin{align}\label{eq:17}
U\left(\alpha, \beta, z\right)
=\frac{1}{\Gamma\left(\alpha \right)}
\int_{0}^{\infty} v^{\alpha-1}\left(1+v\right)^{\beta-\alpha-1}e^{-x v}dv.
\end{align}
\end{linenomath*}
For auxiliary functions $m^{n_{1}}_{n_{2}}\left(p\right)$ thus, we have,
\begin{linenomath*}
\begin{align}\label{eq:18}
m^{n_{1}}_{n_{2}}\left(p\right)
=2^{n_{1}}U\left(n_{2}+1,n_{1}+n_{2}+2,p\right)\Gamma\left(n_{2}+1\right)e^{-p}.
\end{align}
\end{linenomath*}
\section{The JRAF Package}\label{DJRAF}
\subsection{Package overview}
A program code written in a vector form instead of scalar form is performed at the same time on several vector elements so that particularly for modern multi$-$core processors it dramatically improves the performance of code containing loops. The constructed algorithm for computing the Eq. (\ref{eq:13}) is accordingly, optimized based on the vectorization procedure. The relativistic $G^{n_{1},q}_{n_{2},n_{3}}\left(p_{123} \right)$ auxiliary functions are represented in terms of three vectorized sub$-$functions as follows:\\
By re$-$writing the Eq. (\ref{eq:13}) in a simpler form where all the terms in parentheses are expressed with a single function we have,
\begin{linenomath*}
\begin{multline}\label{eq:19}
G^{n_{1},q}_{n_{2},n_{3}}\left(p_{123} \right)\\
=\frac{p_{1}^{n_1}}{\Gamma\left(n_{1}+1 \right)}\frac{1}{2^{2q}}
\sum_{s_{1},s_{2},s_{3}}\left(-1 \right)^{s_{1}+s_{2}+s_{3}}\frac{1}{2^{2s_{2}}}F_{s_{1}}\left(q\right)F_{s_{3}}\left(s_{2}\right)
\\
\times \mathcal{J}^{s_{2},s_{2},0}_{n_{2}+2q-2s_{1}+2s_{2}-2s_{3},n_{3}+2s_{1}+2s_{3}}\left(p_{32}\right),
\end{multline}
\end{linenomath*}
The Eq. (\ref{eq:19}) in a vector form is written as,

\begin{algorithm}\label{j:Gcode}
\SetAlgoLined
  \For{$s_{1}$ $in$ $0$ : $q$}{
    \For{$s_{2}$ $in$ $0$ : $N$}{
    \For{$s_{3}$ $in$ $0$ : $s_{2}$}{
    \For{$s_{4}$ $in$ $0$ : $N$ + $2q$ + $2s_{2}$}{
    $\mathcal{L}_{3}[s_{1},s_{2},s_{3}]$ += $\mathcal{L}_{1}[s_{1},s_{2},s_{3},s_{4}]+ \mathcal{L}_{2}[s_{1},s_{2},s_{3},s_{4}]$
    		}
    $\mathcal{J}_{3}[s_{1},s_{2},s_{3}] = 2^{n_{2}+n_{3}+2q+2s_{2}}\mathcal{B}[s_{1},s_{2},s_{3}]-2^{s_{2}}\mathcal{L}_{3}[s_{1},s_{2},s_{3}]$
	
	$\mathcal{J}_{2}[s_{1},s_{2}]$ += $\left(-1\right)^{s_{3}}
	(1/2^{2s_{2}})F_{s_{3}}\left(s_{2}\right)\mathcal{J}_{3}[s_{1},s_{2},s_{3}]$
    	}
    $\mathcal{J}_{1}[s_{1}]$ += $\left(-1\right)^{s_{1}+s_{2}}F_{s_{1}}\left(q+1\right)\mathcal{J}_{2}[s_{1},s_{2}]$
    }
   $\mathcal{J}$ += $\mathcal{J}_{1}[s_{1}]$
  }
  $G$=$(p_{1}^{n_1}/\Gamma\left(n_{1}+1 \right))(1/2^{2q})\mathcal{J}$
\caption{Method of computation for RAF}
\end{algorithm}
The $\mathcal{B}$ and $\mathcal{L}_{3}$ are three$-$ and four$-$dimensional vectors, respectively. Considering together the Eq. (\ref{eq:13}) and the Algorithm \ref{j:Gcode}, it is easy to clarify which functions are used in the content of these vectors. In order to store elements of $\mathcal{B}$ and $\mathcal{L}_{3}$ vectors, the derived recurrence relationships in previous paper \cite{27_Bagci_2020} make it sufficient instead of three or four to use only two sum indices. The Algorithm \ref{j:Gcode} is accordingly, not for direct computation of the $\mathcal{J}$ vector but to properly re$-$shape and collect all its stored elements. \\
Since they have only one sum indices, the exponential integrals functions $E_{-n}\left(p\right)$ in $\mathcal{B}$ do not require any additional reduction. The assigned vector for Beta functions $B\left[s_{1},s_{2},s_{3}\right]$ with three sum indices, is represented by two sum indices as $B[s_{2},s_{1}+s_{3}]$. It is computed by considering the following initial values and the recurrence relationships,\\
\begin{linenomath*}
\begin{align*}
B\left[1,1 \right]=B_{n_{2}+2q,n_{3}}, \hspace{2mm} B\left[1,2 \right]=B_{n_{2}+2q-2,n_{3}+2}
\\
B\left[2,1 \right]=B_{n_{2}+2q+2,n_{3}}, \hspace{2mm} B\left[2,2 \right]=B_{n_{2}+2q,n_{3}+2},
\end{align*}
\end{linenomath*}
for the row elements, $B\left[s_{2}, 1\right]$, $B\left[s_{2}, 2\right]$,
\begin{linenomath*}
\begin{align}\label{eq:20}
B_{z+2s, z'}
=
\frac{\left(z+2s-1\right) \left(z+2s-2\right)}
{\left(z+z'+2s-1\right)\left(z+z'+2s-2\right)}
B_{z+2s-2,z'},
\end{align}
\end{linenomath*}
then, for the column elements, $B\left[s_{2}, s_{1}+s_{3}\right]$,
\begin{linenomath*}
\begin{align}\label{eq:21}
B_{z-2s,z'+2s}
=\frac{\left(z'+2s-1 \right)\left(z'+2s-2 \right)}{\left(z-2s \right)\left(z-2s+1 \right)}
B_{z-2s+2,z'+2s-2}.
\end{align}
\end{linenomath*}
The elements of $l^{n_{1},q_{1}}_{n_{2}n_{3}}\left(p_{123}\right)$ auxiliary functions  given in the Eq. (\ref{eq:15}) are stored in memory through a series of vectors, the last of which is $\mathcal{L}_{3}[s_{1},s_{2},s_{3}]$. By a minor manipulation as below computationally more efficient implementation is achieved for $l^{n_{1},q_{1}}_{n_{2}n_{3}}\left(p_{123}\right)$,
\begin{linenomath*}
\begin{multline}\label{eq:22}
l^{n_{1},q}_{n_{2}n_{3}}\left(p_{123}\right)
=\frac{p_{1}^{n_{1}}}{\Gamma\left(n_{1}+1\right)}e^{-p_{2}}
\left(-n_{2}\right)_{q}
\sum_{s_{4}}\frac{\left(-n_{2}-q\right)_{s_{4}}}{\left(n_{3}+s_{4}+1\right)s_{4}!}
\\ \times
\big(
m^{n_{2}+q_{1}-s_{4}}_{n_{3}+s_{4}+1}\left(p_{3}\right)
\big/
\left(-n_{2}+s_{4}\right)_{-q}
\big).
\end{multline}
\end{linenomath*}
There are two types of $m^{n_{1}}_{n_{2}}\left(p\right)$ functions and six types of Pochhammer symbols with equal number in each of $\mathcal{L}$ vectors $(\mathcal{L}_{1}, \mathcal{L}_{2})$. The modified form of $l^{n_{1},q_{1}}_{n_{2}n_{3}}\left(p_{123}\right)$ herewith, enables store the elements of vectors for both Pochhammer symbols and $m^{n_{1}}_{n_{2}}\left(p\right)$ auxiliary functions within a same loop. It enables to reduce the number of summation indices from four to two as well. The assigned vectors for computation of Pochammer symbols arise in the Eq. (\ref{eq:22}), from left to right hand$-$side respectively, are called to as $p_{11}, p_{12}, p_{13}$ for $\mathcal{L}_{1}$ and $p_{21}, p_{22}, p_{23}$ for $\mathcal{L}_{2}$. Similarly, the assigned vectors for computation $m^{n_{1}}_{n_{2}}\left(p\right)$ auxiliary functions are called to as $m_{1}$ for $\mathcal{L}_{1}$ and $m_{2}$ for $\mathcal{L}_{2}$. By considering together the Eq. (\ref{eq:15}) and the Eq. (\ref{eq:22}), the recurrence relationships for Pochammer symbols, $m^{n_{1}}_{n_{2}}\left(p\right)$ functions are derived as,
\begin{linenomath*}
\begin{gather*}
p_{11}\left[1,1\right]=p_{11}\left[1,2\right]=p_{12}\left[s_{5},1\right]=p_{13}\left[1,1\right]=p_{13}\left[1,2\right]=1
\\
p_{11}\left[2,1\right]=\left(-n_{2}+2q+2\right)
\hspace{5mm}
p_{11}\left[2,2\right]=\left(-n_{2}+2q+1\right)
\\
p_{12}\left[s_{5},2\right]=-\left(-n_{2}+2q+N-s_{5}\right)
\\
p_{13}\left[2,1\right]=\left(-n_{2}+2q+2\right)
\hspace{5mm}
p_{13}\left[2,2\right]=\left(-n_{2}+2q+3\right).
\end{gather*}
\end{linenomath*}
For the column elements of the $p_{12}\left[s_{5}, s_{6}\right]$ vector,
\begin{linenomath*}
\begin{multline}\label{eq:23}
\left(-\left[n_{2}+2q+N-s_{5}\right]\right)_{s_{6}}
=\left[-\left(n_{2}+2q+N-s_{5}\right)+s_{6}\right]
\\ \times
\left(-\left[n_{2}+2q+N-s_{5}\right]\right)_{s_{6}-1},
\end{multline}
\end{linenomath*}
for the row elements of the $p_{13}\left[s_{2},s_{7}\right]$,
\begin{linenomath*}
\begin{multline}\label{eq:24}
\left(-\left[n_{2}+2p\right]-2s_{2}\right)s_{2}\\
=-\left(\dfrac{n_{2}+2q+2s_{2}}{n_{2}+2q+s_{2}}\right)
\left(-n_{2}-2q-2s_{2}+1\right)
\\ \times
\left(-\left[n_{2}+2q\right]-2s_{2}+2\right)_{s_{2}-1},
\end{multline}
\end{linenomath*}
for the column elements of the $p_{13}\left[s_{2},s_{7}\right]$,
\begin{linenomath*}
\begin{multline}\label{eq:25}
\left(-\left[n_{2}+2q\right]-2s_{2}+s_{7}\right)_{s_{2}}
=-\left(\dfrac{n_{2}+2q+s_{2}-s_{7}+1}{n_{2}+2q+2s_{2}-s_{7}}\right)
\\ \times
\left(-\left[n_{2}+2q\right]-2s_{2}+s_{7}-1\right)_{s_{2}},
\end{multline}
\begin{gather*}
p_{11}\left[s_{2},1\right]=p_{13}\left[s_{2},1\right],
\hspace{2mm}
p_{11}\left[s_{2},2\right]=p_{13}\left[s_{2},2\right]
\\
p_{11}\left[s_{2},s_{5}\right]=p_{13}\left[s_{2},s_{5}\right].
\end{gather*}
\end{linenomath*}
All of the four sum indices explicitly appear in the $m^{n_{1}}_{n_{2}}\left(p\right)$ auxiliary functions. These auxiliary functions involve confluent hyper$-$geometric functions. An efficient approach for accurate calculation of hyper$-$geometric functions with different parameter and variable regimes is still studied in the literature. It herewith would not be advantageous to consider direct use of the Eq. (\ref{eq:18}). The symmetry properties found in previous studies \cite{24_Bagci_2018, 26_Bagci_2018, 27_Bagci_2020} allow avoiding computation of hyper$-$geometric functions and enable the recurrence relations with only two sum indices to be derived. The derived relationships are also consistent with the vectorization procedure followed as a method in the present study.
\begin{linenomath*}
\begin{gather*}
m_{1}\left[1,1\right]=m^{n_{2}+2q}_{n_{3}+1}\left(p_{2} \right)
\hspace{5mm}
m_{1}\left[1,2\right]=m^{n_{2}+2q-1}_{n_{3}+2}\left(p_{2} \right)
\\
m_{1}\left[2,1\right]=m^{n_{2}+2q+1}_{n_{3}+1}\left(p_{2} \right)
\hspace{5mm}
m_{1}\left[2,2\right]=m^{n_{2}+2q}_{n_{3}+2}\left(p_{2} \right)
\end{gather*}
\end{linenomath*}
for the row elements, $m_{1}\left[s_{2}, 1\right]$, $m_{1}\left[s_{2}, 2\right]$,
\begin{linenomath*}
\begin{multline}\label{eq:26}
{m_{1}}^{\left(n_{2}+s_{2}\right)+2q-s_{7}}_{n_{3}+s_{7}+1}\left(p \right)\\
=2\frac{\left[\left(n_{2}+s\right)+n_{3}+2q+p+1 \right]}{p}
{m_{1}}^{\left(n_{2}+s_{2}\right)+2q-\left(s_{7}+1 \right)}_{n_{3}+s_{7}+1}\left(p\right)
\\
+4\frac{\left[s_{7}-\left(n_{2}+s_{2}\right)-2q+1\right]}{p}
{m_{1}}^{\left(n_{2}+s\right)+2q-\left(s_{7}+2\right)}_{n_{3}+s_{7}+1}\left(p \right),
\end{multline}
\end{linenomath*}
for the column elements, $m_{1}\left[s_{2}, s_{7}\right]$,
\begin{linenomath*}
\begin{multline}\label{eq:27}
{m_{1}}^{\left(n_{2}+s_{2}\right)+2q-s_{7}}_{n_{3}+s_{7}+1}\left(p \right)
=\frac{1}{4}\frac{\left(n_{3}+s_{7} \right)}{\left(\left(n_{3}+s_{2}\right)+2q-s_{7}+1 \right]}\\
\times {m_{1}}^{\left(n_{2}+s_{2}\right)+2q-\left(s_{7}-2 \right)}_{n_{3}+s_{7}+1}\left(p\right)
+\frac{1}{2}\frac{\left[\left(n_{2}+s_{2}\right)-n_{3}+2q-2s_{7}-p_{2}+1 \right]}{\left(\left(n_{2}+s_{2}\right)+2q-s_{7}+1\right]} \\
\times {m_{1}}^{\left(n_{2}+s_{2}\right)+2q-\left(s_{7}-1\right)}_{n_{3}+s_{7}}\left(p \right).
\end{multline}
\end{linenomath*}
The elements of vectors $p_{21}, p_{22}, p_{23}$ and $m_{2}$ are obtained by only exchange the indices $n_{2}$ with $n_{3}$. The new indices $s_{5}, s_{6}, s_{7}$ are defined in the Eqs. (\ref{eq:23}$-$\ref{eq:27}) to at first ensure that all the elements of vectors appearing in the Algorithm \ref{j:Gcode} are stored in the memory. The range of these new indices are given as, $0 \leq s_{5} \leq 2N+2q, \hspace{3mm} 0 \leq s_{6} \leq 3N+2q, \hspace{3mm} 0 \leq s_{7} \leq N+4q+4s_{1}.$
\subsection{Installation and usage}
The $Mathematica$ notation is used in the JRAF package for the basic and special mathematical functions, angular momentum coefficients such as Legendre, Laguerre polynomials, hypergeometric functions, spherical harmonics, angular momentum coefficients related with the product of two spherical harmonics located on different centers, Clebsch$-$Gordan and Gaunt coefficients, rotated angular functions ect, so that those familiar with the $Mathematica$ programming language can easily coordinate. These functions are installed one loads the JRAF package by typing \textit{"using JRAF"}. The package is installed via $Julia's$ package manager as,
\begin{lstlisting}[language=C++,basicstyle=\scriptsize]
using Pkg
Pkg.add(path="https://github.com/Nemocas/Nemo.jl.git")
Pkg.add(path="https://github.com/abagciphys/JRAF.jl.git")
\end{lstlisting}
Some additional packages such as $Legendre.jl$ \cite{31_Legendre.jl}, $SphericalHarmonics.jl$ \cite{32_SphericalHarmonics.jl}, $WignerSymbols.jl$ \cite{33_WignerSymbols.jl} are required but only for tests. The $Nemo$ computer algebra package \cite{34_Fieker_2017} is used for accuracy. It is based on $C$ libraries such as $FLINT$,$ANTIC$,$Arb$, $Pari$ and $Singular$. It forms the basis of $JRAF$ through $ccall$, an ordinary function call in $Julia$ \cite{35_Bezanson_2017}. The following syntax for $ccall$ is routinely used in our code,
\begin{lstlisting}[language=C++,basicstyle=\footnotesize]
$ccall((symbol,library),
RetType, (ArgType1,...),Arg1,...)$
\end{lstlisting}
In the Listing \ref{j:MathF}, respectively, we show how to call the Clebsch$-$Gordan and Gaunt coefficients \cite{36_Guseinov_1995, 37_Guseinov_2009}, coefficients related product of two spherical harmonics located on different centers \cite{36_Guseinov_1995} and rotated angular functions \cite{38_Guseinov_2011} as example.
\begin{lstlisting}[caption={Some of angular momentum coefficients available in the JRAF package}, label={j:MathF}, basicstyle=\scriptsize]
$ClebschGordanG(l_1,m_1,l_2,m_2,L,M)$
$GGauntG(l_1,m_1,l_2,m_2,L,M)$
$RotaD(\lambda,l_1,m_1,l_2,m_2,\theta,\phi)$ // for complex spherical harmonics
$Rotad(\lambda,l_1,m_1,l_2,m_2,\theta,\phi)$ // for real spherical harmonics
$SphPCG(q,\alpha,\beta,l_1,\lambda_1,l_2,\lambda_2,\Lambda)$

$\left\lbrace l_1, m_1,l_2,m_2,L,M,\lambda_1,\lambda_2,\lambda,\Lambda\right\rbrace \in Z$
$\left\lbrace \theta, \phi \right\rbrace \in R \vee ArbField$
\end{lstlisting}
See the $math.jl$, $special\_functions.jl$, $angular\_coefficients.jl$ files for entire mathematical expressions and $radial\_coefficients.jl$ for coefficients related with the normalized STO and STSO used in the package for computation for two$-$center one$-$electron overlap, nuclear attraction and kinetic energy integrals over Slater$-$type orbitals with non$-$integer principal quantum numbers. These integrals are included to JRAF package in order to test efficiency of relativistic molecular auxiliary functions. The auxiliary functions via series representation of beta functions are simply called as follows,
\begin{lstlisting}[caption={The relativistic molecular auxiliary functions based on the analytical method}, label={j:MathG},basicstyle=\scriptsize]
$AuxiliaryG(n_1,q,n_2,n_3,p_1,p_2,lim)$ // for $p_3=0$
$AuxiliaryG(n_1,q,n_2,n_3,p_1,p_2,p_3,lim)$ // for $p_3 \ne 0$
$AuxiliaryGr(n_1,q,n_2,n_3,p_1,p_2,p_3,lim)$ // for $p_3 \ne 0$

$\left\lbrace n_1,n_2,n_3,p_1,p_2,p_3,\right\rbrace \in R \vee ArbField$
$\left\lbrace q,lim \right\rbrace \in Z$
\end{lstlisting}
Results for the Eq.(\ref{eq:9}) through the Eq.(\ref{eq:13}) can be obtained with $AuxiliaryG$ (it is found in $gaux\_p123\_bsrep.jl$). In this case, it is advantageous to compute the auxiliary functions separately depending on the parameter $p_{3}$. Results for the Eq.(\ref{eq:9}) through an analytical method based on a recurrence strategy can be obtained with $AuxiliaryGr$ (Plese see \cite{27_Bagci_2020}). Note that, $Cuba$ multidimensional numerical integration library \cite{39_Hahn_2007} is also available to be used by JRAF. The list of functions is found in $cgaux\_p123\_num.jl$. They are defined as follows,
\begin{lstlisting}[caption={The relativistic molecular auxiliary functions based on the numerical integration approximation}, label={j:CubaG},basicstyle=\scriptsize]
$CuhreAuxiliaryG(n_1,q,n_2,n_3,p_1,p_2,lim)$ // for $p_3 \ne 0$
$VegasAuxiliaryG(n_1,q,n_2,n_3,p_1,p_2,p_3,lim)$ // for $p_3 \ne 0$
$SuaveAuxiliaryG(n_1,q,n_2,n_3,p_1,p_2,p_3,lim)$ // for $p_3 \ne 0$

$\left\lbrace n_1,n_2,n_3,p_1,p_2,p_3,\right\rbrace \in R \vee ArbField$
$\left\lbrace q,lim \right\rbrace \in Z$
\end{lstlisting}
Finally, the two$-$center one$-$electron integrals in both lined$-$up and nonlined$-$up molecular coordinate systems can be found in a file which is called as $sto\_mol\_integ\_one\_elect.jl$.
\begin{lstlisting}[caption={The functions for two$-$center one$-$electron overlap integrals in both lined$-$up and nonlined$-$up molecular coordinate systems}, label={j:MolInteg},basicstyle=\scriptsize]
$TwoCenterOverlap(n_1, l_1, n_2, l_2, \lambda, \rho, \tau,lim)$// Analytical
$TwoCenterOverlap(n_1, l_1, m_1,n_2, l_2, m_2,\rho, \tau, \theta, \phi,x, lim)$// Analytical
$CTwoCenterOverlap(n_1, l_1, n_2, l_2, \lambda, \rho, \tau,lim)$// Numerical via Cuhre
$CTwoCenterOverlap(n_1, l_1, m_1,n_2, l_2, m_2,\rho, \tau, \theta, \phi,x, lim)$// Numerical via Cuhre
$VTwoCenterOverlap(n_1, l_1, n_2, l_2, \lambda, \rho, \tau,lim)$// Numerical via Vegas
$VTwoCenterOverlap(n_1, l_1, m_1,n_2, l_2, m_2,\rho, \tau, \theta, \phi,x, lim)$// Numerical via Vegas
$STwoCenterOverlap(n_1, l_1, n_2, l_2, \lambda, \rho, \tau,lim)$// Numerical via Suave
$STwoCenterOverlap(n_1, l_1, m_1,n_2, l_2, m_2,\rho, \tau, \theta, \phi,x, lim)$// Numerical via Suave
$TwoCenterOverlapRec(n_1, l_1, n_2, l_2, \lambda, \rho, \tau,lim)$//RAF recurrence
$TwoCenterOverlapRec(n_1, l_1, m_1,n_2, l_2, m_2,\rho, \tau, \theta, \phi,x, lim)$//RAF recurrence

$\left\lbrace n_1,n_2,n_3,\rho,\tau,\theta,\phi,\right\rbrace \in R \vee ArbField$
$\left\lbrace l_1,m_1,l_2,m_2,\lambda,x\right\rbrace \in Z$
$\left\lbrace lim \right\rbrace \in Z$//for analytical evaluation
$\left\lbrace lim \right\rbrace \in R$//for numerical integration approximation
\end{lstlisting}
The functions for the two$-$center one$-$electron nuclear attraction and kinetic energy integrals are called similarly.
\section{Results and Discussions}\label{ResDis}
In this work a computer program code that is used for computation of relativistic molecular auxiliary functions and their performance on the calculation of two$-$center one$-$electron integrals is presented. The molecular auxiliary functions are used in both solution of the Schr{\"o}dinger and Dirac equations for molecules when the radial part of non$-$integer Slater$-$type orbitals are considered as a basis set. As it is stated in the Section \ref{Intro} Introduction that non$-$analytic near the origin model is considerably different from polynomial model which leads the multi$-$center integrals with non$-$integer power functions to analytically be evaluated via addition theorem. Analytical relations for product of two$-$functions centered on different positions are available. Obtaining compact form expression for multi$-$center integrals with non$-$integer power functions on the other hand, thought to be nearly insurmountable. Studies on various non$-$linear models in physical, chemical and engineering applications are encountered with this problem. So far, it is preferred to either treat empirically or approximate by an analytic function \cite{40_Sombrin_2013}.\\
The molecular integrals over Slater$-$type orbitals arising in the molecular SCF equation are within the scope of such problematic non$-$linear model, yet accurate values for these integrals are achieved through relativistic molecular auxiliary functions and their fully analytical expression in terms of incomplete Beta functions.\\
The efficiency of molecular auxiliary functions are investigated in Tables \ref{gaux_p123}, \ref{overlap}, \ref{nucat} and \ref{KinE}. In these tables, the results for the relativistic molecular auxiliary functions,\\
the overlap,
\begin{multline} \label{eq:28}
	S_{nl m,n'l' m'} \left(\zeta_{a}, \zeta_{b}, \vec{R}_{ab} \right) 
	=\int_{}^{}\chi_{nl m}^{*}\left(\zeta,\vec{r}_{a} \right)\chi_{n'l' m'}\left(\zeta',\vec{r}_{b} \right)dV,
\end{multline}
the nuclear attraction,
\begin{multline} \label{eq:29}
	^{abb}V_{nl m,n'l' m'}\left(\zeta_{a}, \zeta_{b}, \vec{R}_{ab} \right) \\
	=\int_{}^{}\chi_{nl m}^{*}\left(\zeta,\vec{r}_{a} \right)\left(\frac{1}{{r_{b}}}\right) \chi_{n'l' m'}\left(\zeta',\vec{r}_{a} \right)dV,
\end{multline}
and the kinetic energy integrals,
\begin{multline} \label{eq:30}
	T_{nl m,n'l' m'} \left(\zeta_{a}, \zeta_{b}, \vec{R}_{ab} \right)\\
	=\int_{}^{}\chi_{nl m}^{\*}\left(\zeta,\vec{r}_{a} \right)\left(-\frac{1}{2}\nabla^{2} \right)\chi_{n'l' m'}\left(\zeta',\vec{r}_{b} \right)dV.
\end{multline}
are presented depending on upper limit of summation which emerge due to convergent series representation for the molecular auxiliary functions, respectively. The two$-$center kinetic energy integrals are expressed in terms of the overlap integrals using the following simple relationships,
\begin{align}\label{eq:31}
\nabla^{2}Y_{lm}\left(\theta, \varphi\right)
=-\frac{l\left(l+1\right)}{r^2}Y_{lm}\left(\theta, \varphi\right),
\end{align}
\begin{multline}\label{eq:32}
-\frac{1}{2}\nabla^{2}\chi_{nlm}\left(\zeta, \vec{r}\right)\\
=-\frac{1}{2}\zeta^2 \Bigg[
\chi_{nlm}\left(\zeta, \vec{r}\right) \Bigg.
-4\left(\frac{\Gamma\left(2n-1\right)}{\Gamma\left(2n+1\right)} \right)
\chi_{n-1lm}\left(\zeta, \vec{r}\right)\\
\Bigg.
+4\left(n+l\right)\left(n-l-1\right)
\left(\frac{\Gamma\left(2n-3\right)}{\Gamma\left(2n+1\right)} \right)
\chi_{n-2lm}\left(\zeta, \vec{r}\right) \Bigg].
\end{multline}
The solution two$-$center nuclear attraction integrals are derived through the single$-$center potential \cite{9_Bagci_2016},
\begin{multline}\label{eq:33}
	V_{nlm_{l},n'l'm_{l'}}\left(\zeta, \zeta', \vec{R}_{ab}\right)
	=\sum_{LM}\sqrt{\frac{4\pi}{2L+1}}C^{L|M|}(lm,l'm')\\
	\times R^{L}_{nn'}\left(\zeta,\zeta', R_{ab} \right)Y^{*}_{LM}\left(\theta_{R_{ab}}, \vartheta_{R_{ab}} \right).
\end{multline}
Here, the single-center potential $R^{L}_{nn'}\left(\zeta_{A},\zeta'_{A}, R_{AB} \right)$ determined by,
\begin{multline}\label{eq:34}
	R_{n,n'}^{L} \left(\zeta,\zeta', R_{ab} \right)
	= \left(2\overline{\zeta} \right)\Gamma\left [n+n'+L+1 \right]\frac{1}{\left(2\overline{\zeta}R_{ab} \right)^{L+1}} \\
	\times\Bigg\{P \left [n+n'+L+1,2\overline{\zeta}R_{ab} \right] \Bigg. \\+\left. \frac{\left(2\overline{\zeta}R_{ab} \right)^{2L+1}}{\left(n+n'-L\right)_{2L+1}}Q \left [n+n'-L,2\overline{\zeta}R_{ab} \right] \right\},
\end{multline}
with, $C^{L|M|}(lm,l'm')$ are the Gaunt coefficients \cite{36_Guseinov_1995, 37_Guseinov_2009}, $\overline{\zeta}=\zeta+\zeta'$, $Y_{lm}$ are the spherical harmonics, $\vec{R}_{ab}=\vec{r}_{a}-\vec{r}_{b}$, respectively.

The algorithm described for JRAF package in the previous section is also incorporated into a computer program code written in the $Mathematica$ programming language. It is one of the high level symbolic programming language that can handle approximate real numbers with any number of digits. It is suited for modeling a scientific and mathematical problem before a comprehensive analysis because it has close correspondence to mathematical notation. A computer program code that containing loops on the other hand, is more efficient in other languages than the $Mathematica$. The computation time slightly improve if the $\textbf{Table[]}$ functions is used instead of the $\textbf{For[]}$ or $\textbf{Do[]}$ but that still does not allow to compute the molecular auxiliary functions via the Algorithm \ref{j:Gcode}. 
The $Mathematica$ alternatively offers to use $\textbf{Compile[]}$ function. Similarly to machine code of a typical computer, the compiled function is evaluated through an object ($CompiledFunction$). All parameters in an expression are now, numbers (or logical variables). They can be executed quickly but the output limits to $\$MachinePrecision\$$ effective decimal digits (about $16-$digits) \cite{28_Mathematica}. From our experience it is observed that the $\textbf{Compile[]}$ works smoothly for a \say{light} function (a function with a few parameters and loops). The analytical evaluation of relativistic auxiliary functions, on the other hand contain too many parameters and loops in which to produce correct results the $\textbf{Compile[]}$ should more carefully be investigated. Such work provides fast calculation for RAF with $\$MachinePrecision\$$ in the $Mathematica$ programming language. The detail beyond scope of the present study. \\
Two approximations are considered from the numerical integration point of view. The $Cuba$, multidimensional numerical integration library and the Global$-$adaptive method with Gauss$-$Kronrod numerical integration extension. The $Mathematica$ Global$-$adaptive method provides accuracy to arbitrary number of digits, but that costs $CPU$ time. As a supplementary material we have incorporated a $Mathematica$ package (called to as $MRAF$) to our main $JRAF$ package for such comparisons. The $Cuba$ library has four algorithms for multidimensional integration: $vegas$, $suave$, $divonne$ and $cuhre$. These algorithms are significantly faster than $Mathematica$ Global$-$adaptive method but much less accurate. They are available in both $JRAF$ and $MRAF$ packages. They are invoked in almost similar way. The results obtained from these numerical approximations are presented in the Tables \ref{gaux_p123}, \ref{overlap}, \ref{nucat} and \ref{KinE}. In these tables the values in parentheses are $AccuracyGoal$ for the $Mathematica$ Global$-$adaptive method, upper limit of summation for the analytical solution via the series representation of incomplete beta functions (\ref{eq:13}), requested accuracy($N$ means that E$-N$) for Cuba numerical integration algorithm, respectively.\\
In the Table \ref{gaux_p123} the upper limit of summation $\left(N\right)$ for analytical solution is chosen to be 50. It can be seen from this table that with this fixed value of upper limit of summation and given values of parameters the results obtained from the analytical solution for molecular auxiliary functions are accurate at least for 20$-$digits. The $Cuba$ library is performed in both $Julia$ and $Mathematica$ platforms. The computation results are almost identical. They are up to about 10$-$12 digits accurate. They are calculated using the $Cuhre$ multidimensional numerical integration algorithm. \\
Numerical results are assembled in Table \ref{overlap} for overlap integrals in non$-$lined up molecular coordinate system. They are given depending on the upper limit of summation $\left(N\right)$.  Note that, it should be at least $N=70$ to achieve a satisfactory precision. The benchmark values for two$-$center nuclear attraction integrals and two$-$center kinetic energy integrals with non$-$integer principal quantum numbers are first time given in the Tables \ref{nucat}, \ref{KinE}, respectively. Solution for the two$-$center nuclear attraction integrals do not require using the relativistic auxiliary function. It can be derived using only the single$-$center potential, expanded in terms of the non$-$integer Slater$-$type orbitals. In the Table \ref{nucat} thus, the values found in the literature are preferred for comparison. The benchmark results presented in the Table \ref{KinE} for the kinetic energy integrals are compared with $Mathematica$ Global$-$adaptive method. This is because the analytical method derived by the author in \citep{24_Bagci_2018, 26_Bagci_2018, 27_Bagci_2020} recently is the only one so far known to give accurate results. 

The additional computer program code written for the molecular integrals [Eqs. (\ref{eq:28}$-$\ref{eq:30})] is based on the formulas given in \cite{9_Bagci_2016,25_Bagci_2020}. The results show that code presented in this study for the relativistic auxiliary functions based on the analytical method via the series representation of the beta functions allows arbitrary precision floating point calculations.

Dropping the restriction on principal quantum number leads to new features \cite{9_Bagci_2016,25_Bagci_2020, 27_Bagci_2020} (see also references therein) and wide range of application in physics and chemistry. The analytical method used for computation of molecular auxiliary functions is of course open for improvement. The method of computation for the auxiliary functions will be improved, the range of application will be further increased and the $JRAF$ package will be updated accordingly.

\section*{Acknowledgements}
In this study the author A. B. was supported by the Scientific Research Coordination Unit of Pamukkale University under the project number 2020BSP011.
\newpage
\bibliographystyle{}
\bibliography{<your-bib-database>}

\begin{thebibliography}{0}
\bibitem{1_Bagci_2015} A. Ba\u{g}c{\i}, P. E. Hoggan, Phys. Rev. E \textbf{91}(2) (2015) 023303. doi: \url{https://link.aps.org/doi/10.1103/PhysRevE.91.023303}

\bibitem{2_Bagci_2018} A. Ba\u{g}c{\i}, P. E. Hoggan, Rend. Fis. Acc. Lincei \textbf{29}(1) (2018) 191-197. doi: \url{https://doi.org/10.1007/s12210-018-0669-8}

\bibitem{3_Bagci_2018} A. Ba\u{g}c{\i}, P. E. Hoggan and M. Adak, Rend. Fis. Acc. Lincei \textbf{29}(4) (2018) 765-775. doi: \url{https://doi.org/10.1007/s12210-018-0734-3}

\bibitem{4_Bagci_2020} A. Ba\u{g}c{\i}, P. E. Hoggan, Rend. Fis. Acc. Lincei accepted(online) (2020). doi: \url{https://doi.org/10.1007/s12210-020-00953-3}

\bibitem{5_Mathematica} \url{https://www.wolfram.com/mathematica/}

\bibitem{6_Fieker_2017} C. Fieker, W. Hart, T. Hofmann and F. Johansson {\sl Nemo/Hecke: Computer Algebra and Number Theory Packages for the Julia Programming Language} in Proceedings of ISSAC '17 (New York, ACM, 2017) pp. 157-164.

\bibitem{7_Hahn_2007} T. Hahn, Comput. Phys. Commun \textbf{176}(11) (2007) 712-713. doi: \url{https://doi.org/10.1016/j.cpc.2007.03.006}

\end{thebibliography}

\begin{thebibliography}{00}

\bibitem{1_Shabad_2005} A. E. Shabad, J. Phys. A: Math. Gen. \textbf{38}(33) (2005) 7419-7439. doi: \url{https://doi.org/10.1088/0305-4470/38/33/014}

\bibitem{2_Gitman_2013} D. M. Gitman, A. D. Levin, I. V. Tyutin and B. L. Voronov, 
Physica Scripta \textbf{87}(3) (2013) 038104. doi: \url{https://doi.org/10.1088/0031-8949/87/03/038104}

\bibitem{3_Schwerdtfeger_2015} P. Schwerdtfeger, L. F. Pa\v{s}teka, A. Punnett and P. O. Bowman, Nuclear Physics A \textbf{944} (2015) 551-577. doi: \url{https://doi.org/10.1016/j.nuclphysa.2015.02.005}

\bibitem{4_Schwarz_1982} W.H.E. Schwarz, H. Wallmeier, Molecular Physics \textbf{46}(5) (1982) 1045-1061. doi: \url{https://doi.org/10.1080/00268978200101771}

\bibitem{5_Schwarz_1982} W.H.E. Schwarz, E. Wechsel-Trakowski, Chem. Phys. Lett. \textbf{85}(1) (1982) 94-97. doi: \url{https://doi.org/10.1016/0009-2614(82)83468-4}

\bibitem{6_Quiney_1989} H. M. Quiney, I. P. Grant, and S. Wilson, \textit{On the Relativistic Many-Body Perturbation Theory of Atomic and Molecular Electronic Structure}, in Many-Body Methods in Quantum Chemistry. Lecture Notes in Quantum Chemistry, edited by U. Kaldor, Vol. 52, Springer-Verlag, Berlin, 1989, pp. 307–344. doi: \url{https://doi.org/10.1007/978-3-642-93424-7_15}

\bibitem{7_Grant_2007} I. P. Grant, \textit{Relativistic Quantum Theory of Atoms and Molecules}, Springer, New York, 2007. doi: \url{https://doi.org/10.1007/978-0-387-35069-1}

\bibitem{8_Bostock_2011} C. J. Bostock, J. Phys B: At. Mol. Opt. Phys. \textbf{44}(8) (2011) 083001.doi: \url{https://doi.org/10.1088/0953-4075/44/8/083001}

\bibitem{9_Bagci_2016} A. Ba\u{g}c{\i}, P. E. Hoggan, Phys. Rev. E \textbf{94}(1) (2016) 013302. doi: \url{https://link.aps.org/doi/10.1103/PhysRevE.94.013302}

\bibitem{10_Biedenharn_1984} L. C. Biedenharn, J. D. Louck, \textit{Angular Momentum in Quantum Physics},  Cambridge University Press, United Kingdom, 1984. doi: \url{https://doi.org/10.1017/CBO9780511759888}

\bibitem{11_Weniger_2008} E. J. Weniger, J. Phys. A: Math. Theor. \textbf{41}(42), (2008) 425207. doi: \url{https://doi.org/10.1088/1751-8113/41/42/425207}

\bibitem{12_Weniger_2012} E. J. Weniger, J. Math. Chem. \textbf{50}(1), (2012) 17-81. doi: \url{https://doi.org/10.1007/s10910-011-9914-4}

\bibitem{13_Bagci_2014} A. Ba\u{g}c{\i}, P. E. Hoggan, Phys. Rev. E \textbf{89}(5) (2014) 053307. doi: \url{https://link.aps.org/doi/10.1103/PhysRevE.89.053307}

\bibitem{14_Bagci_2015} A. Ba\u{g}c{\i}, P. E. Hoggan, Phys. Rev. E \textbf{91}(2) (2015) 023303. doi: \url{https://link.aps.org/doi/10.1103/PhysRevE.91.023303}

\bibitem{15_Davis_1975} P. J. Davis and P. Rabinowitz, \textit{Methods of Numerical Integration}, Academic Press, New York, 1975, pp. 344-417. doi: \url{https://doi.org/10.1016/B978-0-12-206360-2.50011-X}

\bibitem{16_Mathematica_NInteg} Advanced Numerical Integration in the Wolfram Language\\
\url{https://reference.wolfram.com/language/tutorial/NIntegrateIntegrationStrategies.html}\\
(accessed \today)

\bibitem{17_Weatherford_2005} C. A. Weatherford, E. Red and P. E. Hoggan, Molecular Physics, \textbf{103}(15-16) (2005) 2169-2172. doi: \url{https://doi.org/10.1080/00268970500137261}

\bibitem{18_Weatherford_2006} C. A. Weatherford, E. Red, D. Joseph and P. E. Hoggan, Molecular Physics, \textbf{104}(9) (2006) 1385-1389. \url{https://doi.org/10.1080/00268970500462248}

\bibitem{19_Weatherford_1985} C. A. Weatherford, K. Onda, A. Temkin, Phys. Rev. A \textbf{31}(6) (1985) 3620-3626. doi: \url{https://link.aps.org/doi/10.1103/PhysRevA.31.3620}

\bibitem{20_Gill_2012} A. Gill, J. Segura and T. M. Temme, SIAM J. Sci. Comput. \textbf{34}(6) (2012) A2965-A2981. doi: \url{https://doi.org/10.1137/120872553}

\bibitem{21_Backeljauw_2014} F. Backeljauw, S. Becuwe, A. Cuyt, J. van Deun and W. Lozier, \textbf{90} (2014) 2-20. doi: \url{https://doi.org/10.1016/j.scico.2013.05.006}

\bibitem{22_Jameson_2016} G. J. O. Jameson, The Mathematical Gazette \textbf{100}(548) (2016) 298-306. doi: \url{https://doi.org/10.1017/mag.2016.67}

\bibitem{23_Nemes_2019} G. Nemes, A. B. O. Daalhuls, Math. Comp. \textbf{80} (2019) 1805-1827. doi: \url{https://doi.org/10.1090/mcom/3391}

\bibitem{24_Bagci_2018} A. Ba\u{g}c{\i}, P. E. Hoggan, Rend. Fis. Acc. Lincei \textbf{29}(1) (2018) 191-197. doi: \url{https://doi.org/10.1007/s12210-018-0669-8}

\bibitem{25_Bagci_2020} A. Ba\u{g}c{\i}, Rend. Fis. Acc. Lincei \textbf{31}(2) (2020) 369-385. doi: \url{https://doi.org/10.1007/s12210-020-00899-6}

\bibitem{26_Bagci_2018} A. Ba\u{g}c{\i}, P. E. Hoggan and M. Adak, Rend. Fis. Acc. Lincei \textbf{29}(4) (2018) 765-775. doi: \url{https://doi.org/10.1007/s12210-018-0734-3}

\bibitem{27_Bagci_2020} A. Ba\u{g}c{\i}, P. E. Hoggan, Rend. Fis. Acc. Lincei \textbf{31}(4) (2020) 1089-1103. doi: \url{https://doi.org/10.1007/s12210-020-00953-3}

\bibitem{28_Mathematica} Wolfram Mathematica: Modern Technical Computing\\
\url{https://www.wolfram.com/mathematica/}
(accessed \today)

\bibitem{29_Appell_1925} P. Appell, \textit{Sur les fonctions hyperg\'eom\'etriques de plusieurs variables, les polyn\^omes d'Hermite et autres fonctions sph\'eriques dans l'hyperespace}, Gauthier-Villars, Paris, 1925. url: \url{www.numdam.org/item/MSM_1925__3__1_0/}

\bibitem{30_Abramowitz_1974} M. Abramowitz, I. A. Stegun, \textit{ Handbook of Mathematical Functions with Formulas, Graphs, and Mathematical Tables}, Dover Publica-tions, New York, 1974. doi: \url{https://dl.acm.org/doi/10.5555/1098650}

\bibitem{31_Legendre.jl} Legendre.jl — Calculating Associated Legendre Polynomials\\
\url{https://github.com/jmert/Legendre.jl}
(accessed \today)

\bibitem{32_SphericalHarmonics.jl} T. Limpanuparb, J. Milthorpe, \textit{Associated Legendre Polynomials and Spherical Harmonics Computation for Chemistry Applications}, arXiv: 1410.1748 [physics.chem-ph]. \url{https://arxiv.org/abs/1410.1748}

\bibitem{33_WignerSymbols.jl} Wigner Symbols\\
\url{https://github.com/Jutho/WignerSymbols.jl} (accessed \today)

\bibitem{34_Fieker_2017}   C.  Fieker,  W.  Hart,  T.  Hofmann  and  F.  Johansson, \textit{Nemo/Hecke: Computer Algebra and Number TheoryPackages for the Julia Programming Language} in Pro-ceedings of ISSAC ’17, Association for Computing Machinery (ACM), New York, 2017, pp. 157-164. doi: \url{https://doi.org/10.1145/3087604.3087611}

\bibitem{35_Bezanson_2017} J. Bezanson, A. Edelman, S. Karpinski and V. B. Shah, \textit{Julia:  A Fresh Approach to Numerical Computing}, Siam Rev. \textbf{59}(1), (2017) 65-98. doi: \url{https://doi.org/10.1137/141000671}

\bibitem{36_Guseinov_1995} I. I. Guseinov, J. Mol. Struct. THEOCHEM \textbf{336}(1) (1995) 17-20. doi: \url{https://doi.org/10.1016/0166-1280(94)04101-W}

\bibitem{37_Guseinov_2009} I. I. Guseinov, B. A. Mamedov, E. {\c C}apuro{\u g}lu, J. Theor. Comput. Chem. \textbf{8}(2) (2009) 251-259. doi: \url{https://doi.org/10.1142/S0219633609004782}

\bibitem{38_Guseinov_2011} I. I. Guseinov, J. Math. Chem. \textbf{49}(5) (2011) 1011-1013. doi: \url{https://doi.org/10.1007/s10910-010-9792-1}

\bibitem{39_Hahn_2007} T. Hahn, Comput. Phys. Commun \textbf{176}(11) (2007) 712-713. doi: \url{https://doi.org/10.1016/j.cpc.2007.03.006}

\bibitem{40_Sombrin_2013} J. Sombrin, International Journal of Microwave and Wireless Technologies \textbf{5}(2) (2013) 133-140. doi: \url{https://doi.org/10.1017/S1759078713000172}

\bibitem{41_Mamedov_2004} B. A. Mamedov, Chin. J. Chem. \textbf{22}(6) (2004) 545-548. doi: \url{https://doi.org/10.1002/cjoc.20040220610}

\bibitem{42_Mamedov_2009} B. A. Mamedov, E. \c{C}opuro{\u{g}}lu, MATCH Commun. Math. Comput. Chem. 61 (2009) 553-560. link: \url{https://match.pmf.kg.ac.rs/electronic_versions/Match61/n2/match61n2_553-560.pdf}

\bibitem{43_Guseinov_2002} I. I. Guseinov, B. A. Mamedov, J. Theor. Comput. Chem. \textbf{1}(1) (2002)17-24. doi: \url{https://doi.org/10.1142/S0219633602000130}

\bibitem{44_Mamedov_2012} B. A. Mamedov, E. \c{C}opuro{\u{g}}lu, Appl. Math. Comput. \textbf{218}(15) (2012) 7893-7897. doi: \url{https://doi.org/10.1016/j.amc.2012.01.069}

\end{thebibliography}

\onecolumn
\begin{table}[ht!]
  \centering
\caption{The comparative values of relativistic molecular auxiliary functions}\label{gaux_p123}
\tiny
\resizebox{\columnwidth}{!}{
\begin{threeparttable}
\begin{tabular}{llllllllllll}
\toprule
$n_1$ & $q$ & $n_2$ & $n_3$ & $p_1$ & $p_2$ & $p_3$ & \multicolumn{1}{c}{Results}\\
\midrule
1.1 & 10 & 2.1 & 3.1 & 4.1 & 5.1 & 6.1 &
\begin{tabular}[c]{@{}l@{}}
\hspace{2.5mm}\underline{\textbf{9.69169 58617 01844 36783 81836 63826 82002}} E$+$02 (50)\tnote{a}\\ 
\hspace{2.5mm}\underline{\textbf{9.69169 58617 01844 36783 81836 63}}798 73602 E+02 (50)\tnote{b}\\ 
\hspace{2.5mm}\underline{\textbf{9.69169 58617 0184}}3 73631 52810 48132 44179 E+02 (50)\tnote{c}\\
\hspace{2.5mm}\underline{\textbf{9.69169 58617 018}}5 E+02 (Infinity)\tnote{d}
\end{tabular} \\
\bottomrule
2.1 & 10 & 1.1 & 3.1 & 4.1 & 5.1 & 6.1 &
\begin{tabular}[c]{@{}l@{}}
\hspace{2.5mm}\underline{\textbf{9.43636 92168 48006 16284 17343 75719 57719}} E+02 (50)\tnote{a}\\ 
\hspace{2.5mm}\underline{\textbf{9.43636 92168 48006 16284 17343 75}}681 35383 E+02 (50)\tnote{b}\\ 
\hspace{2.5mm}\underline{\textbf{9.43636 92168 4}}7990 79571 54805 79589 58620 E+02 (50)\tnote{c}\\
\hspace{2.5mm}\underline{\textbf{9.43636 92168 4800}}9 E+02 (Infinity)\tnote{d}
\end{tabular}\\
\bottomrule
3.1 & 10 & 2.1 & 1.1 & 4.1 & 5.1 & 6.1 &
\begin{tabular}[c]{@{}l@{}}
\hspace{2.5mm}\underline{\textbf{1.59485 76412 79536 55278 49760 57498 92476}} E+02 (50)\tnote{a}\\ 
\hspace{2.5mm}\underline{\textbf{1.59485 76412 79536 55278 49760 57}}504 35474 E+02 (50)\tnote{b}\\ 
\hspace{2.5mm}\underline{\textbf{1.59485 76412 795}}41 95537 59700 39698 33024 E+02 (50)\tnote{c}\\
\hspace{2.5mm}\underline{\textbf{1.59485 76412 7953}}98 E+02 (Infinity)\tnote{d}
\end{tabular}\\
\bottomrule
4.1 & 10 & 3.1 & 2.1 & 1.1 & 5.1 & 6.1 &
\begin{tabular}[c]{@{}l@{}}
\hspace{2.5mm}\underline{\textbf{6.05958 87737 18998 83565 42985 28782 66716}} E+00 (50)\tnote{a}\\ 
\hspace{2.5mm}\underline{\textbf{6.05958 87737 18998 83565 42985 287}}98 04857 E+00 (50)\tnote{b}\\ 
\hspace{2.5mm}\underline{\textbf{6.05958 87737}} 19026 73460 92830 01347 72555 E+00 (50)\tnote{c}\\
\hspace{2.5mm}\underline{\textbf{6.05958 87737}} 190095 E+00 (Infinity)\tnote{d}
\end{tabular}\\
\bottomrule
5.1 & 10 & 4.1 & 3.1 & 2.1 & 1.1 & 6.1 &
\begin{tabular}[c]{@{}l@{}}
\hspace{2.5mm}\underline{\textbf{8.50162 73995 71398 04513 17797 38998 36005}} E+14 (50)\tnote{a}\\ 
\hspace{2.5mm}\underline{\textbf{8.50162 73995 71398 0451}}2 35340 82353 96903 E+14 (50)\tnote{b}\\ 
\hspace{2.5mm}\underline{\textbf{8.50162 73995 713}}64 31329 82516 24001 81878 E+14 (50)\tnote{c}\\
\hspace{2.5mm}\underline{\textbf{8.50162 73995 713}}81 E+14 (Infinity)\tnote{d}
\end{tabular}\\
\bottomrule
6.1 & 10 & 5.1 & 4.1 & 3.1 & 2.1 & 1.1 &
\begin{tabular}[c]{@{}l@{}}
\hspace{2.5mm}\underline{\textbf{1.95595 04573 51400 77878 61630 61785 09184}} E+10 (50)\tnote{a}\\ 
\hspace{2.5mm}\underline{\textbf{1.95595 04573 51400 77878 61630 61}}641 47748 E+10 (50)\tnote{b}\\ 
\hspace{2.5mm}\underline{\textbf{1.95595 04573 51}}391 05460 12281 34753 10798 E+10 (50)\tnote{c}\\
\hspace{2.5mm}\underline{\textbf{1.95595 04573 51}}397 E+10 (Infinity)\tnote{d}
\end{tabular}\\
\bottomrule
\end{tabular}
\begin{tablenotes}
\item[a]{$Mathematica$ numerical Global-adaptive method}
\item[b]{Series representation in terms of incomplete Beta functions (Eq. \ref{eq:13})}
\item[c]{$Cuba$ numerical integration algorithm via the $Julia$ programming language}
\item[d]{$Cuba$ numerical integration algorithm via the $Mathematica$ programming language}
\end{tablenotes}
\end{threeparttable}
}
\end{table}
\begin{table}[ht!]
  \centering
\caption{The values of two-center overlap integrals over Slater$-$type orbitals in nonlined$-$up molecular coordinate systems}\label{overlap}
\resizebox{\columnwidth}{!}{
\begin{threeparttable}
\begin{tabular}{llllllllllll}
\toprule
$type$ & $n$ & $l$ & $m$ & $n'$ & $l'$ & $m'$ & $\rho$ & $\tau$ & $\theta$ & $\varphi$ & \multicolumn{1}{c}{res}\\
\midrule
1 & 50.1 & 0 & 0 & 50.0 & 0 & 0 & 5.10 & 0 & 0$^\circ$ & 0$^\circ$ & 
\begin{tabular}[c]{@{}l@{}}
\hspace{2.5mm}\underline{\textbf{9.57914 65146 38189 77903 14416 92566 55702}} E$-$01\tnote{a}\\ 
$-$5.10432 33568 13500 38729 06834 33981 54978 E$+$16 (30)\tnote{b}\\ 
$-$2.70455 22526 89687 52079 97164 43375 18666 E$+$09 (40)\tnote{b}\\ 
\hspace{2.5mm}\underline{\textbf{9.5791}}3 71708 13494 95901 01001 22241 05008 E$-$01 (50)\tnote{b}\\ 
\hspace{2.5mm}\underline{\textbf{9.57914 65146 38189 77}}898 62614 20408 55424 E$-$01 (60)\tnote{b}\\
\hspace{2.5mm}\underline{\textbf{9.57914 65146 38189 77903 14416 925}}56 67193 E$-$01 (70)\tnote{b}\\
\hspace{2.5mm}\underline{\textbf{9.57914 65146 38189 77903 14416 92566 55702}} E$-$01 (80)\tnote{b}\\ 
\hspace{2.5mm}\underline{\textbf{9.57914 65146 38}}211 44252 10226 33627 19002 E$-$01 (50)\tnote{c}\\
\end{tabular} \\
\midrule
1 & 50.1 & 0 & 0 & 50.0 & 0 & 0 & 5.10 & $10^{-6}$ & 0$^\circ$ & 0$^\circ$ & 
\begin{tabular}[c]{@{}l@{}}
\hspace{2.5mm}\underline{\textbf{9.57914 73920 88121 67589 59783 29819 91146}} E$-$01 (50)\tnote{d}\\ 
$-$5.10429 82498 37206 08667 15067 85366 75895 E$+$16 (30)\tnote{b}\\
$-$2.70454 10725 23248 29668 21926 93662 21983 E$+$09 (40)\tnote{b}\\
\hspace{2.5mm}\underline{\textbf{9.5791}}3 80484 77706 30925 99760 61461 93883 E$-$01 (50)\tnote{b}\\
\hspace{2.5mm}\underline{\textbf{9.57914 73920 88121 6758}}5 07987 82376 09265 E$-$01 (60)\tnote{b}\\
\hspace{2.5mm}\underline{\textbf{9.57914 73920 88121 67589 59783 2981}}0 02649 E$-$01 (70)\tnote{b}\\
\hspace{2.5mm}\underline{\textbf{9.57914 73920 88121 67589 59783 29819 91146}} E$-$01 (80)\tnote{b}\\
\hspace{2.5mm}\underline{\textbf{9.57914 73920 881}}50 53757 09690 24516 12353 E$-$01 (50)\tnote{c}\\
\end{tabular}\\
\midrule
1 & 50.0 & 0 & 0 & 50.1 & 1 & -1 & 10.0 & $2/10$ & 10$^{-6\circ}$ & 10$^{-6\circ}$ & 
\begin{tabular}[c]{@{}l@{}}
$-$\underline{\textbf{5.26144 97645 26770 93240 04629 61400 30620}} E$-$17 (50)\tnote{d}\\ 
\hspace{2.5mm}1.02182 38512 81616 09620 88561 63090 97826 E$-$07 (40)\tnote{b}\\ 
$-$\underline{\textbf{5.26144 97645 26770 93240 0}}7525 42183 27960 E$-$17 (60)\tnote{b}\\
$-$\underline{\textbf{5.26144 97645 26770 93240 04629 61400 30620}} E$-$17 (80)\tnote{b}\\
$-$\underline{\textbf{5.26144 97645 267}}56 96648 13695 34144 74157 E$-$17 (50)\tnote{c}\\
\end{tabular}\\
\midrule
1 & 50.3 & 3 & -3 & 50.2 & 2 & 1 & 16.0 & $20/23$ & 30$^{\circ}$ & 30$^{\circ}$ & 
\begin{tabular}[c]{@{}l@{}}
\hspace{2.5mm}\underline{\textbf{2.77810 18374 83364 88927 46276 56594 39028}} E$-$28 (50)\tnote{d}\\ 
\hspace{2.5mm}2.61264 19016 44693 10366 80298 53932 26153 E$-$14 (40)\tnote{b}\\
\hspace{2.5mm}\underline{\textbf{2.77810 18374 83364 8892}}5 72746 98162 53995 E$-$28 (60)\tnote{b}\\
\hspace{2.5mm}\underline{\textbf{2.77810 18374 83364 88927 46276 56594 3902}}0 E$-$28 (80)\tnote{b}\\  
\hspace{2.5mm}\underline{\textbf{2.77810 18374 833}}09 78968 56287 04133 23295 E$-$28 (50)\tnote{c}\\
\end{tabular}\\
\midrule
1 & 50.4 & 4 & -3 & 50.5 & 5 & -4 & 27.0 & $20/45$ & 10$^{\circ}$ & 10$^{\circ}$ & 
\begin{tabular}[c]{@{}l@{}}
\hspace{2.5mm}\underline{\textbf{8.66395 11416 12200 19257 50998 80389 09852}} E$-$04 (50)\tnote{a}\\ 
\hspace{0.5mm}$-$4.06946 77711 02151 27793 04509 57504 79365 E$-$02 (40)\tnote{b}\\
\hspace{2.5mm}\underline{\textbf{8.66395 11416 12200 19257 5099}}2 67155 12351 E$-$04 (60)\tnote{b}\\
\hspace{2.5mm}\underline{\textbf{8.66395 11416 12200 19257 50998 80389 09852}} E$-$04 (80)\tnote{b}\\
\hspace{2.7mm}\underline{\textbf{8.66395 11416}} 06330 64405 25589 40831 78706 E$-$04 (50)\tnote{c}\\
\end{tabular}\\
\midrule
\end{tabular}
\begin{tablenotes}
\item[a]{Ref. \cite{13_Bagci_2014}}
\item[b]{Series representation in terms of incomplete Beta functions (Eq. \ref{eq:13})}
\item[c]{$Cuba$ numerical integration algorithm via the $Julia$ programming language}
\item[d]{$Mathematica$ numerical Global-adaptive method}
\end{tablenotes}
\end{threeparttable}
}
\end{table}
\begin{table}[ht!]
  \centering
\caption{The values of two-center nuclear attraction integrals over Slater$-$type orbitals in nonlined$-$up molecular coordinate systems}\label{nucat}
\resizebox{\columnwidth}{!}{
\begin{threeparttable}
\begin{tabular}{lllllllllllll}
\toprule
$type$ & $n$ & $l$ & $m$ & $\zeta$ & $n'$ & $l'$ & $m'$ & $\zeta'$ & $\theta$ & $\varphi$ & $R$ & \multicolumn{1}{c}{res}\\
\midrule
1 & 3 & 2 & 2 & 12.40 & 3 & 2 & 2 & 10.60 & 0$^\circ$ & 0$^\circ$ & 6.10 & 
\begin{tabular}[c]{@{}l@{}}
\hspace{2.5mm}\underline{\textbf{1.60316 72157 86609 47251 23680 94620 56587}} E$-$01\tnote{a}\\ 
\hspace{2.5mm}\underline{\textbf{1.60316 72157 8661}} E$-$01\tnote{b}\\
\end{tabular} \\
\midrule
1 & 2 & 1 & 0 & 7.60 & 2 & 1 & 1 & 1.50 & 45$^\circ$ & 180$^\circ$ & 2.30 & 
\begin{tabular}[c]{@{}l@{}}
\hspace{2.5mm}\underline{\textbf{2.00987 04344 05018 34280 46705 43909 12301}} E$-$03\tnote{a}\\ 
\hspace{2.5mm}\underline{\textbf{2.00987 043}}38 7478 E$-$03\tnote{b}\\
\end{tabular}\\
\midrule
1 & 2 & 1 & 1 & 6.70 & 2 & 1 & 1 & 4.10 & 135$^\circ$ & 20$^\circ$ & 0.20 & 
\begin{tabular}[c]{@{}l@{}}
\hspace{2.5mm}\underline{\textbf{2.27254 38442 77127 02657 62876 23189 48019}} E$-$00\tnote{a}\\ 
\hspace{2.5mm}\underline{\textbf{2.27254 38442 7713}} E$-$00\tnote{b}\\
\end{tabular}\\
\midrule
1 & 3 & 1 & 0 & 8.60 & 2 & 1 & 1 & 7.40 & 54$^\circ$ & 40$^\circ$ & 4.00 & 
\begin{tabular}[c]{@{}l@{}}
$-$\underline{\textbf{5.42130 98741 10278 47958 74788 89372 12572}} E$-$04\tnote{a}\\ 
$-$\underline{\textbf{5.42130 987}}26 8004 E$-$04\tnote{b}\\
\end{tabular}\\
\midrule
1 & 4 & 3 & 2 & 15.9 & 5 & 3 & 3 & 10.7 & 40$^\circ$ & 30$^\circ$ & 15.5 & 
\begin{tabular}[c]{@{}l@{}}
$-$\underline{\textbf{4.65385 67668 26447 16066 78116 15770 87299}} E$-$06\tnote{a}\\ 
$-$\underline{\textbf{4.65385 67668 26447 16066 78116 1577}} E$-$6\tnote{c}\\
\end{tabular}\\
\midrule
1 & 10 & 9 & -7 & 12.5 & 10 & 8 & -8 & 10.2 & 80$^\circ$ & 240$^\circ$ & 100.7 & 
\begin{tabular}[c]{@{}l@{}}
$-$\underline{\textbf{1.58615 18962 96097 54713 29402 95391 08373}} E$-$06\tnote{a}\\ 
$-$\underline{\textbf{1.58615 18962 96097 54713 29402 9539}} E$-$6\tnote{c}\\
\end{tabular}\\
\midrule
1 & 50 & 31 & -20 & 13.3 & 50 & 31 & 20 & 12.9 & 126$^\circ$ & 320$^\circ$ & 33.0 & 
\begin{tabular}[c]{@{}l@{}}
$-$\underline{\textbf{2.17756 64084 23336 65954 73693 38926 92479}} E$-$42\tnote{a}\\ 
$-$\underline{\textbf{2.17756 64084 23336 65954 73693 3893}} E$-$42\tnote{c}\\
\end{tabular}\\
\midrule
1 & 50 & 31 & -20 & 13.3 & 50 & 31 & 20 & 12.9 & 126$^\circ$ & 320$^\circ$ & 33.0 & 
\begin{tabular}[c]{@{}l@{}}
$-$\underline{\textbf{2.17756 64084 23336 65954 73693 38926 92479}} E$-$42\tnote{a}\\ 
$-$\underline{\textbf{2.17756 64084 23336 65954 73693 3893}} E$-$42\tnote{c}\\
\end{tabular}\\
\midrule
1 & 2.3 & 1 & 1 & 3.70 & 2.5 & 1 & 1 & 2.50 & 120$^\circ$ & 180$^\circ$ & 12.5 & 
\begin{tabular}[c]{@{}l@{}}
\hspace{2.5mm}\underline{\textbf{6.87155 38290 49764 44379 09182 31770 99366}} E$-$02\tnote{a}\\ 
\hspace{2.5mm}\underline{\textbf{6.87155 38290}} 93 E$-$2\tnote{d}\\
\hspace{2.5mm}\underline{6.87155 37746} 8 E$-$02\tnote{e}\\
\end{tabular}\\
\midrule
1 & 6.4 & 5 & 5 & 8.1 & 6.8 & 5 & 4 & 13.8 & 36$^\circ$ & 108$^\circ$ & 8.70 & 
\begin{tabular}[c]{@{}l@{}}
\hspace{2.5mm}\underline{\textbf{2.45289 05630 72333 96494 37330 60231 29083}} E$-$05\tnote{a}\\ 
\hspace{2.5mm}\underline{\textbf{2.45289 0563}}1 23 E$-$5\tnote{d}\\
\hspace{2.5mm}\underline{\textbf{2.45289 05630 7}} E$-$05\tnote{e}\\
\end{tabular}\\
\midrule
1 & 14.6 & 13 & 12 & 21.70 & 13.2 & 11 & 11 & 10.9 & 162$^\circ$ & 288$^\circ$ & 0.03 & 
\begin{tabular}[c]{@{}l@{}}
\hspace{2.5mm}\underline{\textbf{1.21881 82739 66978 02109 80527 89163 70013}} E$-$05\tnote{a}\\ 
\hspace{2.5mm}\underline{\textbf{1.21881 8274}}0 27 E$-$5\tnote{d}\\
\hspace{2.5mm}\underline{\textbf{1.21881 8273}}2 2 E$-$05\tnote{e}\\
\end{tabular}\\
\midrule
1 & 20.6 & 18 & 15 & 13.80 & 25.6 & 16 & 14 & 9.50 & 20$^\circ$ & 60$^\circ$ & 14.30 & 
\begin{tabular}[c]{@{}l@{}}
$-$\underline{\textbf{1.15016 84587 27269 63624 02716 36736 45953}} E$-$05\tnote{a}\\ 
$-$\underline{\textbf{1.15016 84587 2}}8 E$-$5\tnote{d}\\
$-$\underline{\textbf{1.15017}} 1141 06 E$-$05\tnote{e}\\
\end{tabular}\\
\bottomrule
\end{tabular}
\begin{tablenotes}
\item[a]{Eq.\ref{eq:13})}
\item[b]{Ref. \cite{41_Mamedov_2004}}
\item[c]{Ref. \cite{42_Mamedov_2009}}
\item[d]{Ref. \cite{43_Guseinov_2002}}
\item[e]{Ref. \cite{44_Mamedov_2012}}
\end{tablenotes}
\end{threeparttable}
}
\end{table}

\begin{table}[ht!]
  \centering
\caption{The values of two-center kinetic energy integrals over Slater$-$type orbitals in nonlined$-$up molecular coordinate systems}\label{KinE}
\resizebox{\columnwidth}{!}{
\begin{threeparttable}
\begin{tabular}{lllllllllllll}
\toprule
$type$ & $n$ & $l$ & $m$ & $\zeta$ & $n'$ & $l'$ & $m'$ & $\zeta'$ & $\theta$ & $\varphi$ & $R$ & \multicolumn{1}{c}{Results}\\
\midrule
1 & 1 & 0 & 0 & 1.186 & 1 & 0 & 0 & 1.186 & 90$^\circ$ & 30$^\circ$ & 3.987 &
\begin{tabular}[c]{@{}l@{}}
$-$\underline{\textbf{1.07207 65660 76439 40215 59332 79505 56130}} E$-$02 (50)\tnote{a}\\ 
$-$\underline{\textbf{1.07207 65660 76439 40215 59332 79505 56130}} E$-$02 (80)\tnote{b}\\ 
$-$\underline{\textbf{1.07207 65660 7}}4842 01283 38921 43031 49166 E$-$02 (50)\tnote{c}\\ 
\end{tabular} \\
\midrule
1 & 1 & 0 & 0 & 1.186 & 2 & 1 & 1 & 1.30 & 90$^\circ$ & 60$^\circ$ & 2.30 &
\begin{tabular}[c]{@{}l@{}}
\hspace{2.5mm}\underline{\textbf{1.15661 13009 97389 64535 81162 75999 84072}} E$-$01 (50)\tnote{a}\\ 
\hspace{2.5mm}\underline{\textbf{1.15661 13009 97389 64535 81162 75999 84072}} E$-$01 (80)\tnote{b}\\ 
\hspace{2.5mm}\underline{\textbf{1.15661 13009 97}}461 23292 16521 15937 81041 E$-$01 (50)\tnote{c}\\
\end{tabular}\\
\midrule
1 & 1 & 0 & 0 & 1.186 & 2 & 1 & -1 & 1.30 & 90$^\circ$ & 60$^\circ$ & 2.30 &
\begin{tabular}[c]{@{}l@{}}
$-$\underline{\textbf{2.00330 95379 35818 54671 06227 20618 33457}} E$-$01 (50)\tnote{a}\\ 
$-$\underline{\textbf{2.00330 95379 35818 54671 06227 20618 33457}} E$-$01 (80)\tnote{b}\\ 
$-$\underline{\textbf{2.00330 95379 35}}942 54000 78568 60495 71228 E$-$01 (50)\tnote{c}\\
\end{tabular}\\
\midrule
1 & 2.3 & 1 & 1 & 3.70 & 2.5 & 1 & 1 & 2.50 & 120$^\circ$ & 180$^\circ$ & 1.25 &
\begin{tabular}[c]{@{}l@{}}
\hspace{2.5mm}\underline{\textbf{3.66326 74787 88443 63447 78669 22664 05272}} E$-$10 (50)\tnote{a}\\ 
\hspace{2.5mm}\underline{\textbf{3.66326 74787 88443 63447 78669 22664 05272}} E$-$10 (80)\tnote{b}\\ 
\hspace{2.5mm}\underline{\textbf{3.6}}5084 01829 15509 78924 66313 30922 48778 E$-$10 (50)\tnote{c}\\
\end{tabular}\\
\midrule
1 & 4.1 & 3 & -3 & 3.70 & 4.1 & 2 & 1 & 2.50 & 30$^\circ$ & 30$^\circ$ & 1.25 &
\begin{tabular}[c]{@{}l@{}}
\hspace{2.5mm}\underline{\textbf{2.71140 81788 45886 25040 50731 46689 43726}} E$-$09 (50)\tnote{a}\\ 
\hspace{2.5mm}\underline{\textbf{2.71140 81788 45886 25040 50731 46689 43726}} E$-$09 (80)\tnote{b}\\ 
\hspace{2.5mm}\underline{\textbf{2.71}}895 99482 65152 73280 82593 65885 12258 E$-$09 (50)\tnote{c}\\
\end{tabular}\\
\midrule
1 & 7.5 & 4 & -3 & 3.70 & 8.5 & 5 & -4 & 2.50 & 10$^\circ$ & 10$^\circ$ & 1.25 &
\begin{tabular}[c]{@{}l@{}}
\hspace{2.5mm}\underline{\textbf{3.87720 87457 41408 48188 62888 17585 50270}} E$-$07 (50)\tnote{a}\\ 
\hspace{2.5mm}\underline{\textbf{3.87720 87457 41408 48188 62888 17585 50270}} E$-$07 (80)\tnote{b}\\ 
\hspace{2.5mm}\underline{\textbf{3.8}}6118 82223 84108 35612 97071 46519 19932 E$-$07 (50)\tnote{c}\\
\end{tabular}\\
\bottomrule
\end{tabular}
\begin{tablenotes}
\item[a]{$Mathematica$ numerical Global-adaptive method}
\item[b]{Series representation in terms of incomplete beta functions (Eq. \ref{eq:13})}
\item[c]{$Cuba$ numerical integration algorithm via the $Julia$ programming language}
\end{tablenotes}
\end{threeparttable}
}
\end{table}

\end{document}